# Impact of Annealing Temperature on the Energy Storage Performance of $CoO_2$ Nanoparticles Synthesized via Solid State Reaction

Shoaib Akram[a], Muneeb ur Rahman[b], Fazli Maula[a,b], Osama Tariq Satti[c,d], Shahbaz Afzal[e], Fawad Ali[c,f*]

[a]Department of Physics, University of Punjab, Lahore, Pakistan.
[b]National Institute of Vacuum Science and Technology, NCP, Islamabad, Pakistan.
[c]Department of Chemistry, Abdul Wali Khan University, Mardan, Pakistan
[d]School of Manufacture Science and Engineering, Southwest University of Science and Technology, China
[e]Department of Physics, University of Education Lahore, DG Khan Campus, 32200, Pakistan.
[f]Nanophotonics Research Centre, Shenzhen University and Key Laboratory of Optoelectronic Devices and Systems of Ministry of Education and Guangdong Province, College of Optoelectronics Engineering, Shenzhen University, Shenzhen 518060, China.

*Corresponding author: Fawad Ali, Email: 2451493004@email.szu.edu.cn; fawadali2022@163.com; http://orcid.org/0000-0002-6528-4070.*

**Research Highlights**

- ❖ The solid-state reaction was used to synthesize $CoO_2$ nanostructured material.
- ❖ For the unannealed material, the energy bandgap of $CoO_2$ is 2.00 eV, whereas for the annealed material it ranges from 1.77 to 1.86 eV.
- ❖ The calculated specific capacitances of $CoO_2$, $CoO_2$ (250°C), $CoO_2$ (300°C) nanostructured material at scan rates of (10) mVs$^{-1}$ is (223, 348, and 473) Fg$^{-1}$.

**Abstract**

The solid-state reaction was used to synthesize $CoO_2$ nanostructured material. Cobalt nitrate tetrahydrate (Co(NO$_3$)$_2$·4H$_2$O) and sodium oxide (NaOH) were combined to produce $CoO_2$ nanostructured material. The three synthesized working electrodes were each tested individually, using 3 M KOH as the electrolyte. The CV analysis of a three-electrode system revealed redox peaks, indicating Faradaic processes. The estimated specific capacitances of $CoO_2$, $CoO_2$ (250°C), $CoO_2$ (300°C) nanostructured material at scan rates of (10) mVs$^{-1}$ is (223,

348, and 473) Fg$^{-1}$. The diffraction peaks at 2θ = 26.264º, 33.527º, 37.579, 51.264 and 54.367º correspond respectively to the diffraction planes of 111, 112, 200, 211, and 311 of $CoO_2$ nanostructured material. The annealing temperature, which affects the bandgap, can influence the size, shape, and crystallinity of nanostructures. For the unannealed material, the energy bandgap of $CoO_2$ is 2.00 eV, whereas for the annealed material it ranges from 1.77 to 1.86 eV.

## 1. Introduction

Environmental pollution is a significant result of using non-renewable resources like fossil fuels, which are becoming more expensive each year as reserves diminish [1], [2]. It is crucial to prioritize the development of sustainable green energy, such as wind and solar power [3]–[5]. Advancements in energy storage technology have enhanced the storage and transportation of electricity from renewable sources. Rechargeable batteries and supercapacitors have traditionally been the primary choices for storing chemical energy. Rechargeable lithium-ion batteries, widely used today, provide high voltage, energy density, and excellent safety [6]–[8]. As the demand for lithium-ion batteries rises, the availability of lithium resources is decreasing rapidly. Due to its abundance and low cost, sodium, an alkali metal, has been receiving increasing attention in recent years. Sodium-ion batteries continue to face challenges due to their inadequate cycle performance [9]–[11]. Super capacitors (SCs) provide faster charging and discharging, higher power density, longer lifespan, and safer operation in contrast to rechargeable batteries. Unfortunately, SCs suffer from low energy density. Overcoming low energy density in SCs often involves developing high-performance electrode materials. Energy storage systems have become indispensable in today's world because of the proliferation of mobile electronic devices, electric vehicles, and new energy vehicles [12]–[15]. Electrochemical containers, known as SCs, are gaining attention for their safe operation, cycle performance, charging capacity, and power density [16].

Transition metal materials, especially oxides, are widely preferred for electrode materials in SCs. Cobalt-based materials are advantageous for SCs due to their abundance, stable cycles, electroactive sites, high capacitance, and strong conductivity. $Co_3O_4$ and other cobalt-based materials have undergone extensive research, resulting in notable progress. Cobalt oxide nanoparticles found in various nanostructures display special properties including semi-conductivity, piezoelectricity, and optical features [17], [18]. Cobalt oxide nanoparticles are being evaluated for various uses like Nano sensors, energy storage, cosmetics, nano-electronic, and nano-optics.

Cobalt (II) and ammonium oxalate were reacted by Manteghi et al [10] in their study to generate a cobalt oxalate complex. To manage the particle size, a surfactant like CTAB and F-127 was utilized. To obtain nano cobalt oxide, the calcined precipitate was characterized using FTIR, SEM, TEM, and XRD techniques. The pure and nano-sized particles had an average size smaller than 40 nm. The $Co_3O_4$@Ni foam electrode ($Co_3O_4$@NF) attained a peak capacitance of 351 F g$^{-1}$ in a 2 M KOH solution with a scan rate of 0.85 A g$^{-1}$. The electrode shows exceptional long-term stability, retaining almost 98.6% of its initial specific capacitance after 1000 testing cycles.

Velhal et al [16] successfully grew various nanostructures of cobalt oxide on Ni foam via surface-modifying and a one-pot hydrothermal synthesis process. The Co3O4 took on various shapes based on whether Triton X-100, CTAB, or its original state was used as a mediator. The CTAB-induced dandelion flower nanonetwork exhibits a specific capacitance of 521.63 Fg$^{-1}$, as revealed by electrochemical measurements. The electrode we created maintains excellent electrochemical stability for over 3000 charge-discharge cycles interconnected network, resulting in high electronic conductivity and low diffusion resistance. Its initial capacitance remains at 95.29% even after 3000 charge-discharge cycles. Electrode charge storage in supercapacitors was improved through surface modification of $Co_3O_4$ material.

The structure of cobalt oxide nanostructured materials comprises tiny cobalt oxide particles. Their applications are diverse and include batteries, supercapacitors, and electro-catalysis. They find applications in gas sensors, magnetic substances, and materials used in photocatalysis.

In this study, we synthesized cobalt oxide nanostructured materials to improve the physical properties of metal oxide nanoparticles for energy storage applications. Various characterization tools will analyze the Cyclic voltammetry, optical, structural, surface micrograph, and elemental composition of the synthesized material.

## 2. Experimental Section
### 2.1. Materials

The chemicals for the synthesis of $CoO_2$ nanostructured material are; 1.5 g of Cobalt nitrate tetrahydrate $Co(NO_3)_2·4H_2O$ and 1.2 g of sodium oxide (NaOH). All chemicals obtained from Sigma Aldrich have a 99% purity level.

### 2.2. Methodology

The solid-state reaction was used to synthesize $CoO_2$ nanostructured material. Cobalt nitrate tetrahydrate ($Co(NO_3)_2·4H_2O$) and sodium oxide (NaOH) were combined to produce $CoO_2$ nanostructured material. The chemicals were mixed with the salts and ground for 15 minutes, repeating the process twice. The initial calcination took 17 hours at a temperature of 950°C, with a steady temperature increase of 10°C per minute. To complete the second calcination, grind for 25 minutes twice and then heat at 10oC/min for 17 hours at 950°C. The powder was milled for two cycles of 25 minutes each before being heated to 1200°C for 12 hours, increasing the temperature by 10°C per minute. Two separate annealing processes were conducted on the $CoO_2$ nanostructured material at 250 and 300°C for 12 hours each. The Gamry Potentiostat and Galvanostat use specific electrodes, including a $CoO_2$ nanostructured material working electrode, a Platinum counter electrode, and an Ag/AgCl reference electrode. The Gamry Reference 3000, a Potentiostat/Galvanostat/ZRA with 32 V/1.5, was used for all electrochemical measurements at room temperature. The electrolyte consisted of a 3 M NaCl concentration. This procedure is followed to adequately prepare the working electrode. Mix CoO2 nanostructured material, carbon black, and PVDF binder in NMP solvent to create a slurry with a 90:5:5 ratio. The slurry was applied to Nickel foam and allowed to dry overnight in an oven to enhance the bond between the active material and the current collector. The newly created electrode was employed for optical analysis and electrochemical measurements, such as cyclic voltammetry (CV). The crystal structures and orientations of both irradiated and unirradiated nanoparticles were examined using an X-ray diffractometer. We conducted our investigation in the 2theta range from 15 to 60°. SEM was used to analyze the morphology and elemental compositions. The optical properties of these materials were examined using a UV-visible spectrophotometer within the range of 300 to 1000 nm.

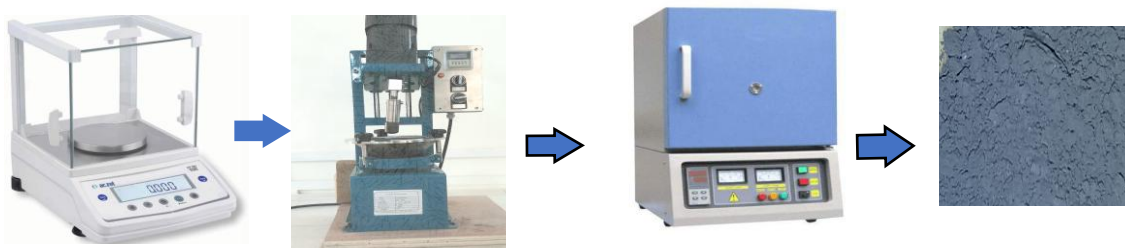

**Figure 1:** Synthesis stages of $CoO_2$ nanostructured material.

**3. Results and Discussion**

## 3.1. Cyclic Voltammetry

Cyclic Voltammetry (CV) was used to assess the electrochemical characteristics of CoO2 nanostructured material, specifically to evaluate its super-capacitance properties. A three-electrode system, including an Ag/AgCl reference electrode, a platinum counter electrode, and a working electrode on nickel foam, was employed for the electrochemical investigations. The three synthesized working electrodes were each tested individually using 3 M KOH as the electrolyte. By analyzing the scan rate, the electrodes' capacitance performance was evaluated using CV. Notable peaks and a strong pseudo-capacitive potential were observed in the three electrodes, which were scanned at a range of 10 mVs$^{-1}$. By analyzing a CV plot and utilizing Equation (1) [19], [20], it becomes feasible to differentiate between various capacitances.

$$C_{SP} = \frac{1}{2mk\Delta V} \int IdV \qquad (1)$$

Equation 1 denotes the specific capacitance as ($C_{SP}$), m denotes the active mass loaded on the electrode, k is the scan rate in mVs$^{-1}$, and $\Delta V$ is the potential window while $\int IdV$ is the integral of area under the CV curve. The CV analysis of a three-electrode system revealed redox peaks, indicating the presence of Faradaic processes. The estimated specific capacitances of $CoO_2$, $CoO_2$ (250°C), $CoO_2$ (300°C) nanostructured material at scan rates of (10) mVs$^{-1}$ is (223, 348, and 473) Fg$^{-1}$. The material's pseudo-capacitive properties become apparent when its potential is measured and charges can rapidly transfer in both directions during Faradaic reactions $CoO_2$ nanostructured material annealed at 300°C demonstrated the highest specific capacitance.

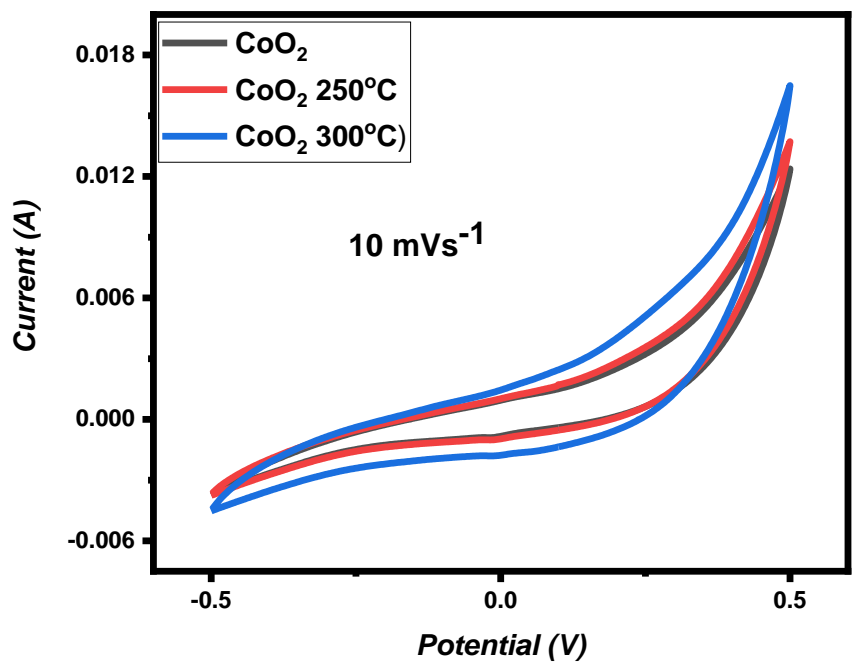

**Figure 2:** Redox behavior of $CoO_2$ Nanoparticles at different annealing temperatures.

## 3.2. Structural Properties

The XRD study reveals the crystal structures of $CoO_2$ nanostructured material as presented in Figure 3. The diffraction value of $CoO_2$ nanostructured material at $2\theta = 54.367°$ confirmed the characteristic peak of $CoO_2$ nanostructured material. The diffraction peaks at $2\theta = 26.264°$, $33.527°$, $37.579$, $51.264$ and $54.367°$ correspond respectively to the diffraction planes of 111, 112, 200, 211, and 311 of $CoO_2$ nanostructured material. The crystallite size of the electrode material was calculated using equation 1[21]–[29]

$$D = \frac{k\lambda}{\beta \cos\theta} \qquad (1)$$

Where k is 0.9 and λ is 0.12406 nm (wavelength of X-ray source) and β is FWHM in radians. Table 1 shows the calculation of the crystallite size of $CoO_2$ nanostructured material. Altering the temperature between 250 and 300°C causes structural changes in $CoO_2$ nanostructured material during annealing. This process influences the material's properties and performance. The $CoO_2$ nanostructured material's crystallite size grows as the annealing temperature rises. This suggests an improvement in the crystallinity and growth of the material's grains. By annealing $CoO_2$ nanostructured material at varying temperatures, the crystallite is improved. The presence of these crystallites impacts both the material and optical bandgap.

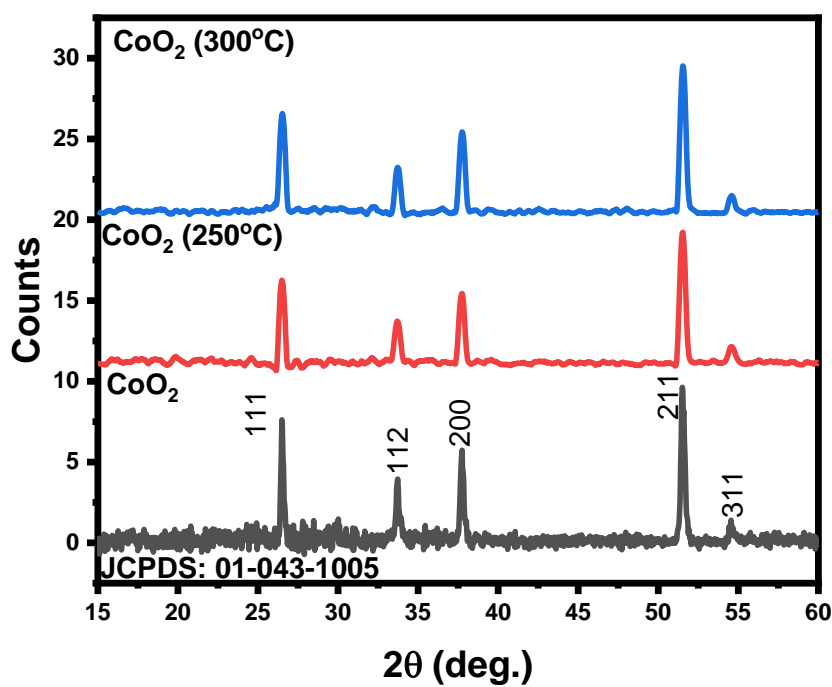

**Figure 3:** XRD pattern of CoO$_2$ nanostructured material.

**Table 1:** structural properties of CoO$_2$ nanostructured material.

| Material | 2 theta (deg.) | D (spacing) | Å | β | hkl | D (nm) | σ (line/m$^2$) X 10$^{15}$ |
|---|---|---|---|---|---|---|---|
| CoO$_2$ | 26.264 | 3.392 | 5.876 | 0.756 | 111 | 1.884 | 8.575 |
| | 33.527 | 2.672 | 5.344 | 0.758 | 112 | 1.911 | 8.333 |
| | 37.579 | 2.393 | 4.786 | 0.759 | 200 | 1.931 | 8.168 |
| | 51.264 | 1.781 | 3.984 | 0.762 | 211 | 2.019 | 7.466 |
| | 54.367 | 1.687 | 4.132 | 0.766 | 311 | 2.036 | 7.345 |
| CoO$_2$ (250°C) | 26.768 | 3.329 | 5.767 | 0.843 | 111 | 1.692 | 1.064 |
| | 33.982 | 2.637 | 5.275 | 0.848 | 112 | 1.711 | 1.040 |
| | 38.043 | 2.364 | 4.729 | 0.849 | 200 | 1.728 | 1.019 |
| | 51.916 | 1.760 | 3.937 | 0.851 | 211 | 1.813 | 9.261 |
| | 54.782 | 1.675 | 4.103 | 0.853 | 311 | 1.832 | 9.074 |
| CoO$_2$ (300°C) | 26.768 | 3.329 | 5.767 | 0.864 | 111 | 1.650 | 1.117 |
| | 33.982 | 2.637 | 5.275 | 0.866 | 112 | 1.675 | 1.085 |

| | | | | | | |
|---|---|---|---|---|---|---|
| 38.043 | 2.364 | 4.729 | 0.869 | 200 | 1.689 | 1.067 |
| 51.916 | 1.760 | 3.937 | 0.872 | 211 | 1.769 | 9.724 |
| 54.782 | 1.675 | 4.103 | 0.878 | 311 | 1.780 | 9.614 |

## 3.3. Surface Morphology

Figure 4 reveals a distinctive surface morphology of the $CoO_2$ nanostructured material. The high surface area and porous structure of this material attribute to its enhanced electrochemical performance. The surface micrograph heavily influenced the behavior of $CoO_2$ nanostructured material in various applications. The nano-ball within a micrograph of $CoO_2$ nanostructured material influences the capacitance and energy storage capacity of supercapacitors. Expanding the interlayer spacing of $CoO_2$ nanostructured material can enhance energy storage capabilities. The scalability of ultra-capacitors relies on the criticality of the surface of $CoO_2$ nanostructured material. The potential of $CoO_2$ nanostructured material for high-power and energy density is attributed to its high conductivity and specific surface area. $CoO_2$ nanostructured material is a superb option for supercapacitor electrodes. The $CoO_2$ nanostructured material exhibits a porous nano-ball surface structure, suggesting oxide in the Co electrode lattice. Including oxide in the Co lattice structure enhances the surface micrograph of the Co electrode, resulting in improved energy storage capabilities. Figure 5 displays the elemental analysis of $CoO_2$, including all the elements found in the materials.

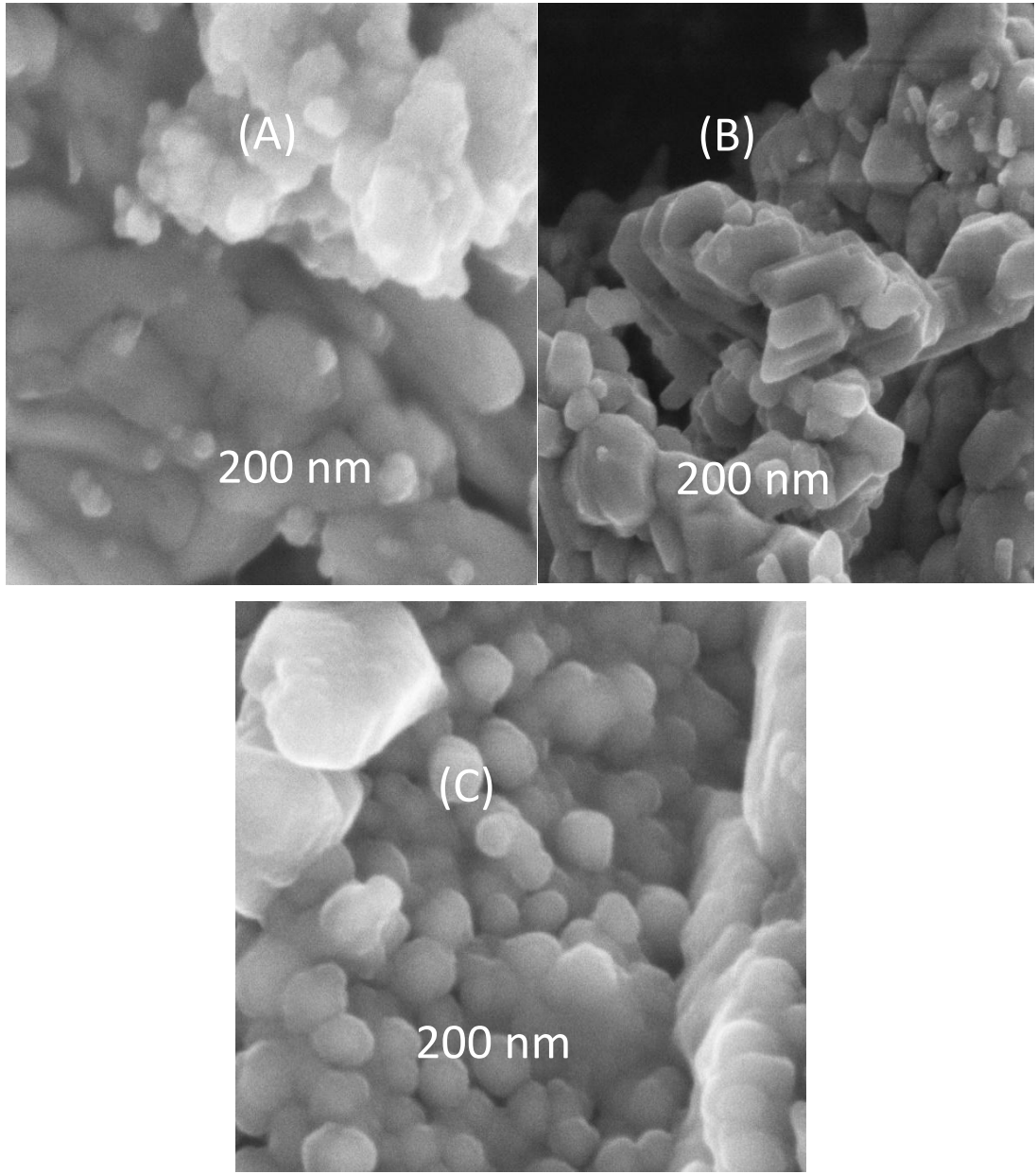

**Figure 4:** SEM of CoO$_2$ nanostructured material (A) Unannealed, (B) 250°C, and (C) 300°C.

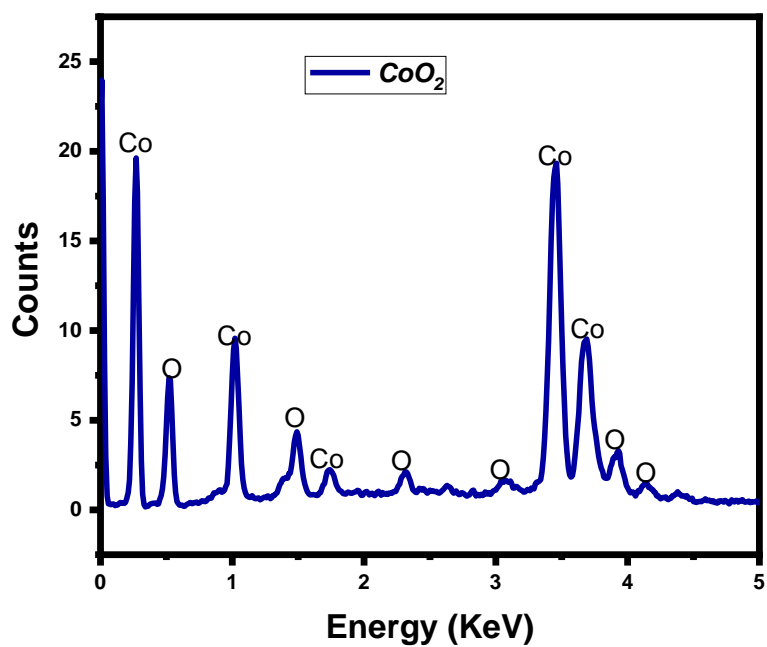

**Figure 5: EDXs of CoO$_2$ nanostructured material.**

### 3.4. Optical Properties

Figure 6 (a) shows the absorbance properties of CoO2. The highest absorbance occurs in the CoO2 material that has not been annealed, as it absorbs more energy in the visible region of the spectra. The absorbance of the annealed material decreases with increasing annealing temperature. CoO2 nanostructured materials exhibit different absorbance properties depending on annealing temperatures of 250 and 300°C. Alterations in the crystal structure, surface morphology, and optical bandgap of the material cause the changes in absorbance. The annealing temperature significantly impacts the absorbance of CoO$_2$ nanostructures. Increased annealing temperatures improve crystallinity, decrease defects, and enhance optical properties. This leads to increased absorbance in the visible and near-infrared regions. Tailored absorbance properties make CoO$_2$ nanostructures useful in photocatalysis, energy storage, and sensing. Their efficient light absorption makes them ideal for solar energy conversion, batteries, and optoelectronic devices. Figure 6 (b) shows the transmittance properties of CoO$_2$. The CoO$_2$ material that hasn't been annealed absorbs more energy in the visible region, which leads to the lowest transmittance. The transmittance of the annealed material increases as the annealing temperature increases. The transmittance of CoO$_2$ nanostructured materials is affected by the annealing temperature. Increased transmittance is typically observed with higher annealing temperatures as a result of improved crystallinity and reduced defects. As the annealing temperature rises, the size of $CoO_2$ nanostructures also increases. Grain growth increases and grain boundaries decrease due to higher temperatures. By controlling the annealing temperature, one can adjust the optical properties of cobalt oxide nanostructures, including transmittance and absorption. This enables the creation of materials with tailor-made optical properties for different uses. Figure 6 (c) illustrates the reflectance properties of $CoO_2$. Unannealed $CoO_2$ material has higher energy absorption in the visible region, resulting in maximum reflectance. The reflectance of $CoO_2$ nanostructured material changes with annealing temperature (250°C to 300°C) due to alterations in its structural and optical properties. The reflectance of cobalt oxide nanostructured material is greatly affected by the annealing temperature. Increased annealing temperatures typically result in greater crystallinity and grain growth, leading to improved reflectance. The $CoO_2$ nanostructured material undergoes structural and optical changes during the annealing process. The material's reflectance properties are affected by changes like the formation of crystal phases and the modification of the bandgap. Figure 6 (d) illustrates the bandgap energy of $CoO_2$. The investigation focused on the energy bandgap of $CoO_2$ nanostructured materials synthesized at various annealing temperatures ranging from 250 to 300ºC. The annealing temperature affects the energy bandgap of $CoO_2$ nanostructures. The bandgap energy decreases as the annealing temperature increases. The structural and morphological properties of $CoO_2$ nanostructures closely influence their energy bandgap. The annealing temperature, which affects the bandgap, can influence the size, shape, and crystallinity of nanostructures. For the unannealed material, the energy bandgap of $CoO_2$ is 2.00 eV, whereas for the annealed material it ranges from 1.77 to 1.86 eV.

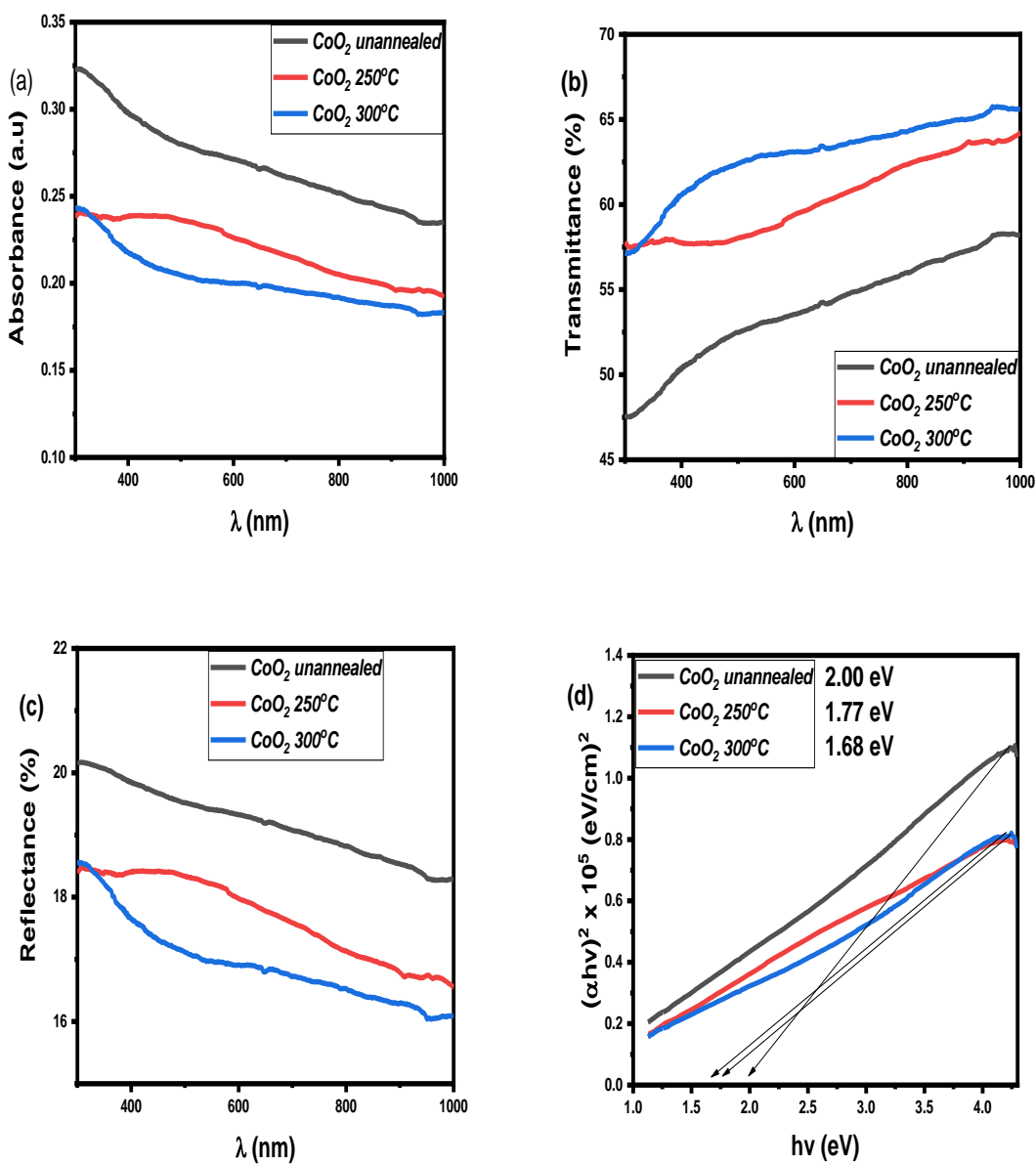

**Figure 6:** UV-visible of $CoO_2$ nanostructured material (a) absorption spectra, (b) transmission spectra, (c) reflection spectra, and (d) bandgap energy spectra.

## 4. Conclusions

We have successfully used the solid-state reaction technique to synthesize $CoO_2$ nanostructured material. The three synthesized working electrodes were each tested individually, using 3 M KOH as the electrolyte. The absorbance of the annealed material decreases with increasing annealing temperature. CoO2 nanostructured materials exhibit different absorbance properties depending on annealing temperatures of 250 and 300°C. Alterations in the crystal structure, surface morphology, and optical bandgap of the material cause the changes in absorbance. The CV analysis of a three-electrode system revealed redox peaks, showing Faradaic processes. The estimated specific capacitances of $CoO_2$, $CoO_2$ (250°C), $CoO_2$ (300°C) nanostructured material at scan rates of (10) $mVs^{-1}$ is (223, 348, and 473) $Fg^{-1}$. The diffraction peaks at $2\theta =$ 26.264°, 33.527°, 37.579, 51.264 and 54.367° correspond respectively to the diffraction planes of 111, 112, 200, 211, and 311 of $CoO_2$ nanostructured material. The annealing temperature, which affects the bandgap, can influence the size, shape, and crystallinity of nanostructures. For the unannealed material, the energy bandgap of $CoO_2$ is 2.00 eV, whereas for the annealed material it ranges from 1.77 to 1.86 eV.

**Availability of Data**

Upon request, data is available.

**Declaration of competing interest**

The authors declare that they have no personal or financial conflicts that could have influenced the research described in this paper.